\documentclass[letterpaper]{mandc2019}
\usepackage{tabls}
\usepackage{cites}
\usepackage{epsf}
\usepackage{appendix}
\usepackage{ragged2e}
\usepackage[top=1in, bottom=1.in, left=1.in, right=1.in]{geometry}
\usepackage{enumitem}
\setlist[itemize]{leftmargin=*}
\usepackage{caption}
\usepackage{color}
\usepackage[capitalize]{cleveref}
\usepackage{amsbsy}
\usepackage{amsmath}
\usepackage{gensymb}
\usepackage{subcaption}
\usepackage{float}
\usepackage{tikz}
\usepackage[makeroom]{cancel} 
\usepackage{multirow} 
\usepackage{booktabs} 
\usepackage{lscape}
\usepackage{bm}

\title{TOWARDS A MULTIPHYSICS MODEL FOR TUMOR RESPONSE \\
TO COMBINED-HYPERTHERMIA-RADIOTHERAPY TREATMENT}
\author{%
  \textbf{Japan K. Patel$^1$, Richard Vasques$^1$, and Barry D. Ganapol$^2$} \\
  $^1$The Ohio State University, Department of Mechanical and Aerospace Engineering\\
201 W.~19$^{\text{th}}$ Avenue, Columbus, OH 43210 \vspace{5pt}\\ 
  $^2$The University of Arizona, Department of Aerospace and Mechanical Engineering \\
    1130 N Mountain Avenue, Tucson, AZ 85721 \\ 
\\ 
  \url{patel.3545@osu.edu}, \url{vasques.4@osu.edu}, \url{ganapol@cowboy.arizona.edu}
}
\newcommand{\authorHead}      
           {J.~K.~Patel, R.~Vasques, B.~D.~Ganapol}  
\newcommand{\shortTitle}      
           {Multiphysics Model for Tumor Response to Combined-Hyperthermia-Radiotherapy Treatment}  
\begin{document}
\maketitle
\justify 

\begin{abstract}
We develop a multiphysics-based model to predict the response of localized tumors to combined-hyperthermia-radiotherapy (CHR) treatment. 
This procedure combines hyperthermia (tumor heating) with standard radiotherapy to improve efficacy of the overall treatment.
In addition to directly killing tumor cells, tumor heating amends several parameters within the tumor microenvironment. 
This leads to radiosensitization, which improves the performance of radiotherapy while reducing the side-effects of excess radiation in the surrounding normal tissue.
Existing tools to model this kind of treatment consider each of the physics separately.  
The model presented in this paper accounts for the synergy between hyperthermia and radiotherapy providing a more realistic and holistic approach to simulate CHR treatment. 
Our model couples radiation transport and heat-transfer with cell population dynamics. 
\end{abstract}

\section{INTRODUCTION} \label{sec1}

Radiotherapy is one of the most widely used modalities for palliative and therapeutic cancer care \cite{steelbook}. 
Its performance depends on energy deposition, and the resulting effects on tumor and surrounding healthy tissue. 
While the dose profile is controlled through the choice of radiation and delivery method \cite{radonc}, its effects on tissue can be manipulated using chemical radiosensitizers \cite{lehnert} and/or tumor heating \cite{hyperrt}. 
The use of radiosensitizers enhances cell kill and allows for optimization of radiation dosage and scheduling such that the collateral damage to the normal tissue can be minimized.
Cytotoxic effects of heat shocks and their role in radiosensitization for radiotherapy have been of interest since the 1970s.
\vspace{-5pt}

There are mechanistic differences in how heat and radiation damage cells \cite{radbio}.
This allows for additive use of hyperthermia and radiotherapy. 
The role of hyperthermia in radiochemosensitization, without the introduction of systemic toxicity, is also well known \cite{peeken}.
This allows for the synergistic use of hyperthermia and radiotherapy.
Therefore, hyperthermia can be used with radiotherapy both additively and synergistically.
Several randomized trials have demonstrated clinical benefits of using hyperthermia in combination with radiotherapy \cite{radbio}.
Hyperthermia is a fully approved part of cancer care in European countries like Germany and the Netherlands.
\vspace{-5pt}

Radiation dosimetry has traditionally been treated as single physics \cite{attix}.
However, the modeling of tumor response to radiotherapy could benefit from the coupling of dose deposition with other relevant physics. 
When patients are exposed to heat and radiation, tumors undergo constitutive changes. 
Taking these changes into account simultaneously with the radiation dose deposition, should provide a more realistic and effective approach to model treatment. 
For example, hyperthermia increases radiosensitivity in certain types of tumor cells, making tumors more susceptible to radiation treatment \cite{hyperrt}.
Therapies such as Combined-Hyperthermia-Radiotherapy (CHR) make use of this effect.\vspace{-5pt} 

To appropriately model the overall effect, the synergy between features such as tumor heating, radiation dosimetry, and tumor dynamics should be considered simultaneously. 
Furthermore, there are cases in which the target organs are in constant motion (e.g.~lungs), thereby constantly shifting the location of the tumor (in center-of-mass frame). 
This motion should be accounted for in real time while treating tumors using external radiation beams and mathematically modeling their response. 
Coupled/multiphysics modeling could prove to be a powerful tool to determine tumor dynamics in such systems. 
Other treatment modalities include a combination of radiotherapy with immunotherapy \cite{radimm}, and radiotherapy with chemotherapy \cite{radche}. 
For such combined treatments, the accurate determination of tumor response should also be performed using multiphysics models.
\vspace{-5pt}

Several studies have presented work on mathematical tumor dynamics modeling. 
Ordinary and partial differential equations are used to model these systems in \cite{altrock} and \cite{radmichor}. 
Radiotherapy-specific tumor dynamics models have been introduced in \cite{rockne} and \cite{sachs}. 
Moreover, internal radiation dosimetry has been extensively treated using the Monte Carlo method \cite{rogers}.
In this work, we introduce the idea of a novel multiphysics-based approach to mathematically model CHR systems. 
This model couples radiation transport (and dosimetry), heat transfer, and cell population dynamics.
To our knowledge, this is the first attempt to develop a multiphysics model for tumor response to CHR.
\vspace{-5pt}

This is by no means an exhaustive study.
The test problems in this preliminary work \textit{do not} model real-world treatment scenarios; the governing equations and treatment specifics are heavily simplified.
These are not the subject of investigation in this report; our goal is to provide a first step towards building a fully-coupled multiphysics model for this type of treatment, with the focus of this paper being the introduction to the coupled approach.
\vspace{-5pt}

The remainder of this paper is organized as follows. 
In \cref{sec:gov_eq}, we introduce relevant physics, governing equations, coupling, and numerical approach. 
In \cref{sec3}, we furnish our example problems and provide numerical results. 
We conclude this paper with a brief discussion in \cref{sec4}.
\vspace{-5pt}

\section{GOVERNING EQUATIONS} 
\label{sec:gov_eq}

The model to describe tumor dynamics during CHR treatment incorporates three physics: 1) radiation transport to model internal dosimetry, 2) heat transfer to incorporate the effects of tumor heating, and 3) cell population dynamics to determine how the tumor responds to the combination of hyperthermia with radiotherapy. 
Our preliminary model is one-dimensional in space, monoenergetic, and assumes isotropic scattering.
\vspace{-5pt}

\subsection{Radiation Transport and Dosimetry}\vspace{-5pt}

The following monoenergetic Boltzmann transport equation with isotropic scattering is used to model the flux distribution \cite{lm}:
\begin{equation}
\label{transport1}
\frac{1}{v}\frac{\partial \psi}{\partial t} + \mu\frac{\partial \psi}{\partial x} + \sigma_t(x) \psi =  \frac{\sigma_{s}(x)}{2} \phi(x, t) + \frac{Q(x, t)}{2},
\end{equation}
where $\psi = \psi(x,\mu, t)$ represents the angular flux at position $x$ and along direction $\mu$ at time $t$; $v$ is the photon speed; $\phi$ is the scalar flux; $\sigma_t$ is the total macroscopic cross-section; $ \sigma_{s}$ is the $0^{th}$ moment of the scattering cross-section; and $Q$ is an internal source.
The dose is calculated using the following equation \cite{martin}:
\begin{subequations}
\begin{equation}
\label{dose1}
D(x, t) = \left(\frac{\mu_{en}}{\rho}\right)\Phi(x, t) E,
\end{equation}
\begin{equation}
\label{fluence}
\Phi(x, t) = \int_{0}^{t} dt' \phi(x, t').
\end{equation}
\end{subequations}
Here, $D$ represents the dose; $\frac{\mu_{en}}{\rho}$ is the mass energy absorption coefficient; $E$ is the energy per photon; and $\Phi$ is the fluence at time $t$. 
We calculate the effective dose $D_E$ using the standard linear-quadratic model \cite{radbio}:
\begin{equation}
\label{edose1}
D_E(x, t) = \alpha D(x, t) + \beta D^2(x,t),
\end{equation}
where $\alpha$ and $\beta$ are radiobiology parameters that determine relative contribution of each term in the sum toward the total radiation effect \cite{radbio}. 
The cell survival probability $S$ is chosen in such a way that a larger dose results in smaller survival probability, as given by \cite{rockne}:
\begin{equation}
\label{survival}
S(x,t) = e^{-D_E(x, t)}.
\end{equation}

\subsection{Hyperthermia and Heat Transfer}\vspace{-5pt}

Thermal medicine is slowly increasing its presence in cancer care. Several techniques, such as radiofrequency and ultrasound ablation, have been developed to provide focused heating inside tumors \cite{radbio}.
While the use of a more sophisticated heat transport model is warranted, we use simple heat conduction to model hyperthermia \cite{cannon}:
\begin{equation}
\label{heat}
\rho c_p \frac{\partial T(x, t)}{\partial t} = \frac{\partial }{\partial x} \kappa \frac{\partial T(x, t)}{\partial x} + q(x),
\end{equation}
where $c_p$, $\rho$, and $\kappa$ are the specific heat capacity, the material density, and the thermal conductivity of tissue respectively (assumed constant).
Moreover, $q$ is the volumetric heat source.
\vspace{-5pt}

\subsection{Cell Population Dynamics}\vspace{-5pt}

We assume that tumor growth depends on three factors: proliferation, invasion, and cell kill induced by treatment. 
Proliferation determines the balance between cell division and cell loss due to natural death; cell proliferation is higher in tumors. 
Invasion essentially determines the transport of tumor cells in the area of interest; continuum-based approaches to model tumor dynamics often use Fick's law to describe invasion. 
Two proliferation models are widely considered in the literature: exponential and logistic. 
Though less suitable for problems considering realistic tumor growth time-frames (months to years), we use the exponential model because our test problems consider tumor dynamics over a short period of thirty minutes. 
The following balance equation determines tumor cell concentration, $c(x,t)$, over time \cite{rockne}:
\begin{equation}
\label{td1}
\frac{\partial c}{\partial t} = \frac{\partial}{\partial x}\left(I\frac{\partial c}{\partial x}\right) + pc(x,t) - R(x, t)c(x,t).
\end{equation}
The equation above equates the rate of change of tumor cell concentration with the sum of net dispersal rate, net proliferation rate, and loss of tumor cells due to therapy.
Here, $I$ is the diffusion coefficient representing motility of tumor cells; $p$ is the proliferation rate; and $R$ represents the effect of therapy on cell kill \cite{rockne,harpold}.
Under stand-alone radiotherapy, 
\begin{equation}
\label{rrt}
R=1-S.
\end{equation}
Patient-specific invasion and proliferation rates can be obtained using MRI images \cite{rockne}.
We define how supplemental heat affects radiotherapy next.
\vspace{-5pt}

\subsubsection{Radiosensitivity parameter} \vspace{-5pt}

As mentioned in \cref{sec1}, heat affects the radiosensitivity of tissue. 
Traditionally, this is expressed using the thermal enhancement ratio (TER) \cite{radbio}, which defines the ratio of dose required to produce a given level of biological damage with and without heat. 
Experiments have measured TER for various types of tumors \cite{gillette}. 
Some experiments have seen TER values as high as $4.3$ under favorable environments \cite{radbio}. 
In this paper, we express this in a different way: we introduce a radiosensitivity parameter, $\xi$, to collectively account for the additive and synergistic effects of hyperthermia.
We define this parameter as the ratio of the biological damage caused by a given amount of radiation dose to tissue with and without heating. 
Setting the relevant biological effect to be $R$, 
\begin{equation}
\label{ter}
\xi(T_{avg}, t) = \frac{R_{CHR}}{R_{RT}},
\end{equation}
where $R_{CHR}$ quantifies the combined effect of heat and radiation on irradiated tissue, and $R_{RT}$ represents the effect of stand-alone radiotherapy.
$T_{avg}$ is the time-averaged value of temperature at time $t$.
Combining \cref{rrt,ter}, we obtain
\begin{equation}
\label{rchr}
R_{CHR} = \xi R_{RT} = \xi(1-S).
\end{equation}
At normal body temperature, $\xi =1$ irrespective of time; we also set $\xi$ at time zero to be unity.
We arbitrarily assume $\xi=2.5$ for our system when the tumor is heated to an average temperature of $45 \degree$C over $30$ minutes.
In order to define $\xi$ elsewhere, we employ bilinear interpolation \cite{nm} between the data points $\xi(37, 0) = 1$, $\xi(37, 30)=1$, $\xi(45, 0)=1$  and $\xi(45, 30)=2.5$. This returns:
\begin{equation}
\label{rsp}
\xi(T_{avg},t) = 1 - 0.23125T_{avg} + 0.00625(T_{avg}\times t).
\end{equation}
Several simplifying assumptions have been made in this model. 
We do not consider temperature dependent cross-sections for transport.
Material density change with heating is ignored.
Moreover, we assume that the decay heat is negligible compared to the constant volumetric heat source in the heat equation.
These simplifications allow us to get a one-way coupled system.
We plan to eliminate these assumptions and model this system with a fully-coupled framework in a later paper.
\vspace{-5pt}

\section{ONE-WAY COUPLED CHR MODELING}
\label{sec3}

The interaction of \cref{transport1,dose1,edose1,survival,td1,heat,rrt,rchr,rsp} determines the tumor response over time.
While our problem is set in slab geometry and is one-way coupled, real-world problems are in general more complex. 
Transport alone can depend on seven independent variables \cite{lm}, making its full coupling with other physics difficult. 
Moment-based nonlinear acceleration schemes, such as nonlinear diffusion acceleration (NDA), aid in accelerating the convergence of slowly converging physics and allow us to isolate the angular flux from the coupled set of physics by introducing a discretely consistent low-order (LO) equation  \cite{patelFBR}. 
The system of equations for nonlinear diffusion acceleration is the following:
\begin{subequations}
\begin{equation}
\label{ho1}
\frac{1}{v}\frac{\partial \psi^{HO}}{\partial t} + \mu\frac{\partial \psi^{HO}}{\partial x} + \sigma_t \psi^{HO} =  \frac{\sigma_{s}(x)}{2} \phi^{LO} + \frac{Q}{2},
\end{equation}
\begin{equation}
\label{lo}
\frac{1}{v}\frac{\partial \phi^{LO}}{\partial t} + \frac{\partial}{\partial x}\left(-D\frac{\partial \phi_{LO}}{\partial x} + \hat{D}\phi^{LO}\right) + \sigma_a \phi^{LO} = Q,
\end{equation}
\begin{equation}
\label{c1}
\hat{D} = \frac{J^{HO} + D\frac{\partial \phi^{HO}}{\partial x}}{\phi^{HO}},
\end{equation}
\begin{equation}
\label{c2}
J^{HO} = \int_{-1}^1 \mu \psi d\mu\,.
\end{equation}
\end{subequations}
Here, $J$ is the current, $D = \frac{1}{3 \sigma_t}$ is the diffusion coefficient, and $\hat{D}$ is the consistency parameter. 
The high-order (HO) transport equation (\ref{ho1}) couples with the diffusion-like LO equation (\ref{lo}) via the closure relations described in \cref{c1,c2}, forming the high-order low-order (HOLO) system of equations.
Nonlinear diffusion acceleration is achieved by simultaneous solution of \cref{ho1,lo,c1,c2}.
We discretize the HO equation using the standard backward Euler method in time and diamond-difference/discrete ordinates scheme in space/angle.
The heat, cell population dynamics, and the LO equations employ backward Euler and central finite difference for time/space discretization.
\vspace{-5pt}

Although the schemes used in this paper are well-known, we briefly present them for completeness.
The discretized transport equation is written as:
\begin{subequations}
\begin{equation}
\label{rtdisc}
\frac{1}{v}\frac{\psi_{i,n,k}-\psi_{i,n,k-1}}{\Delta t} + \mu_n \frac{\psi_{i_{1+\frac{1}{2}}, n, k} -\psi_{i_{1-\frac{1}{2}}, n, k}}{\Delta x} + \sigma_{t, i}\psi_{i,n,k} = \frac{\sigma_{s, i}}{2}\phi_{i, k} + \frac{Q_{i,k}}{2}
\end{equation}
\begin{equation}
\label{closure1}
\psi_{i, n, k} = \frac{\psi_{i_{1+\frac{1}{2}}, n, k}+\psi_{i_{1-\frac{1}{2}}, n, k}}{2}.
\end{equation}
\end{subequations}
Here, $i$, $n$, and $k$ represent spatial, angular, and temporal indices respectively. 
Integer index values represent cell-centers and fraction values represent cell-edges. 
\Cref{dose1,edose1,survival} can be easily discretized in the same fashion. 
The LO equation uses the following discretization \cite{knollJFNK}:
\begin{align}
\label{lodisc}
\frac{1}{v} \frac{\phi_{i,k}-\phi_{i,k-1}}{\Delta t} -D \frac{\phi_{i+1, k} -2\phi_{i, k}+ \phi_{i+1, k}}{\Delta x^2} +&\frac{\hat{D}_{i+\frac{1}{2}}}{2 \Delta x}(\phi_{i+1,k} + \phi_{i, k}) \\
& - \frac{\hat{D}_{i-\frac{1}{2}}}{2 \Delta x}(\phi_{i,k} + \phi_{i-1,k}) + \sigma_a \phi_{i,k}  = Q_{i,k}.\nonumber
\end{align}
The standard centered-space-backward-time discretization for the heat equation is given by
\begin{equation}
\label{heatdisc}
\rho c_p \frac{T_{i, k} - T_{i, k-1}}{\Delta t} = \kappa \frac{T_{i+1, k} - 2T_{i, k} + T_{i-1, k}}{\Delta x^2} + q_{i},
\end{equation}
while the cell population dynamics equation is written as
\begin{equation}
\label{tddisc}
\frac{c_{i, k} - c_{i, k-1}}{\Delta t} = I \frac{c_{i+1, k} - 2C_{i, k} + c_{i-1, k}}{\Delta x^2} + p c_{i, k} -R_{i,k}c_{i,k}.
\end{equation}
The coupling and solution can be represented via the following Newton step:
\begin{equation}
\left[\begin{array}{cccccc}
J_{\phi^{LO} \phi^{LO}} & 0 & 0 & 0 & 0 & 0 \\
J_{\phi^{LO}\Phi} & J_{\Phi\Phi} & 0 & 0 & 0 & 0 \\
0 & J_{\Phi D_E} & J_{D_E D_E} & 0 & 0 & 0  \\
0 & 0 & 0 & J_{T T} & 0 & 0 \\
0 & 0 & J_{D_E R} & J_{T R} & J_{R R} & 0\\
0 & 0 & 0 & 0 & J_{R c} & J_{c c}
\end{array}\right] \left[\begin{array}{c}
\delta \phi^{LO} \\
\delta \Phi \\
\delta D_E \\
\delta T \\
\delta R \\
\delta c
\end{array}\right] = -\left[\begin{array}{c}
F_{\phi^{LO}} \\
F_{\Phi} \\
F_{D_E} \\
F_T \\
F_R \\
F_c
\end{array} \right],
\end{equation} 
where $F_{\phi^{LO}}$,$F_{\Phi}$, $F_{D_E}$, $F_T$, $F_R$, and $F_c$ are the residual forms of the relevant equations for flux, fluence, effective dose, temperature, effect of therapy on cell kill, and tumor cell concentration. 
Since the Jacobian matrix above is not  block-diagonal, the physics are interrelated and therefore warrant a multiphysics treatment. 
The one-way coupling (represented by the lower-triangular Jacobian matrix) allows us to solve the problem in a serial fashion, in which we solve transport first to determine flux, fluence, dose, effective dose, and the survival probability (in that order).
Next, we solve the heat equation to determine the temperature, followed by heat-adjusted effect of radiation on tissue.
Finally, we calculate the tumor cell concentration using parameters obtained by solving the previous equations and move on to the next time step.
This kind of modeling allows us to continuously track tumor cell density over time.
For each of the test problems, we choose a uniform time step of $\Delta t = 0.025$ $s$ and $270$ spatial cells.
The solutions obtained are converged to at least four digits precision for this discretization.
We set the angular discretization for the transport equation to eight angles. 
The convergence tolerance for transport solves is set to $10^{-6}$. 
The heat and tumor dynamics equations are solved using MATLAB's backslash function \cite{matlab}.
\vspace{-5pt}

\subsection{Test Problem}\vspace{-5pt}

We consider the evolution of tumor-cell concentration distribution in a three-region slab over a period of thirty minutes. 
The first and the third regions, each $14$ mm thick, represent healthy soft-tissue.
The second soft-tissue region (in the middle) is $2$ mm thick, and has uniformly distributed tumor cells with a concentration of $2^{20}$ $\frac{cells}{cm^3}$.
This region maintains an exponentially decaying internal radiation source with a half life of $75$ days.  
The average photon energy of the source is $0.4$ MeV.
In order to induce hyperthermia, this region also has a constant volumetric heat source.
Specific values of the radiation and the heat source will be provided in the next subsection.
We use vacuum boundaries for transport, and open boundaries for the heat and tumor dynamics equations. 
We assume the parameters presented in \cref{tab1} \cite{rockne,warrell,kaeri}.
We follow existing literature and assume similar continuum properties for both healthy and tumor tissue.
We represent the transport in healthy and tumor tissues using the same parameters. 
The total cross-section (total attenuation coefficient) is $0.1053578$ $cm^{-1}$; and the scattering cross-section is $0.105324$ $cm^{-1}$ \cite{kaeri}.
\begin{table}[!htb]
  \centering
  \caption{\bf{Parameters}}
\begin{tabular}{ |c|c|c|c|c|c|c|c|c|}\hline\label{tab1}
 \textbf{I}  & \textbf{p} & \pmb{$\alpha$}  & \pmb{$\frac{\alpha}{\beta}$}  & $\bm\kappa$ & \pmb{$c_p$} & \pmb{$\rho$}  & \pmb{$\frac{\mu_{en}}{\rho}$} \\
  $\mathrm{[mm/yr]}$ & [/yr]  & [kg/J]&[$\mathrm{J/kg}$] &[W/m.K] &[kJ/kg.K] &[g/cc] & [$\mathrm{cm^2/g}$]  \\
  \hline
  4.29 & 35.13 & 0.203 & 10 & 0.51 & 3.68 & 1.0 & 0.0325\\ \hline 
\end{tabular}
\end{table}

\subsection{Numerical Results}\vspace{-5pt}

First, we demonstrate that the use of a scale-bridging HOLO algorithm returns the same tumor response profile to CHR treatment as stand-alone source iteration \cite{lm}. We also analyze how increasing radiation source strength affects the tumor response.
Specifically, we consider three initial source strengths for the tumor region: $Q_0 = 1.4 \times 10^5$, $2.8 \times 10^5$, and $5.6 \times 10^5$ $\frac{photons}{cm^3 s}$.
Everything else is the same as previously described.\vspace{-10pt}
\begin{figure}[bht]
\centering
\begin{subfigure}{.48\textwidth}
  \centering
  \includegraphics[width=1\linewidth]{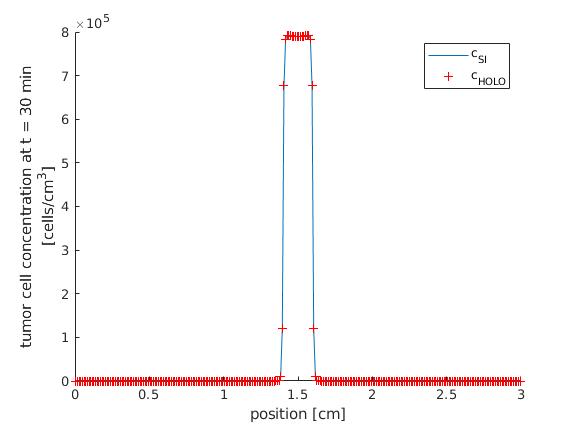}
  \caption{HOLO vs. Source Iteration}
  \label{fig:sub1}
\end{subfigure}
\begin{subfigure}{.48\textwidth}
  \centering
  \includegraphics[width=1\linewidth]{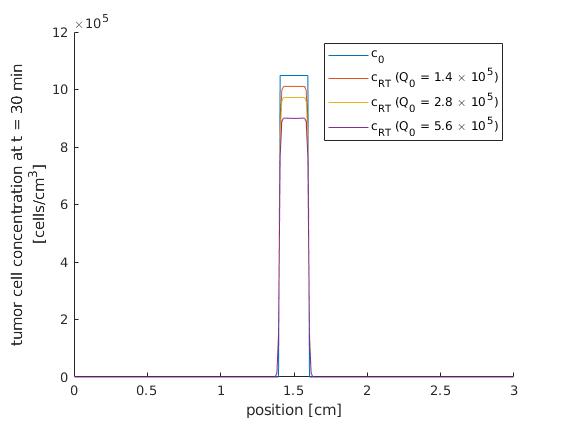}
  \caption{Tumor response to different sources}
  \label{fig:sub2}
\end{subfigure}
\caption{Tumor response to stand-alone radiotherapy}
\label{fig:test}
\end{figure}
\vspace{-20pt}

We observe in \cref{fig:test} that both methods return tumor cell concentration within the prescribed tolerance.
This means it is valid to introduce HOLO acceleration into the multiphysics algorithm.
Moreover, according to \cref{fig:sub2}, increasing source strength improves the overall cell kill as expected.
We note that this study is not prescriptive.
\vspace{-5pt}

Next, we consider tumor dynamics in three scenarios: 1) untreated tissue (no radiation or heat source), 2) slab with an initial isotropic internal radiation source of $5.6\times10^{5}$ $\frac{photons}{cm^3 s}$ in the tumor region, and 3) slab with an initial isotropic internal radiation source of $5.6\times10^{5}$ $\frac{photons}{cm^3 s}$ and a heat source of $\frac{1}{3}$ $\frac{W}{cm^3}$ in the tumor region. 
\vspace{-5pt}

\begin{figure}[h]
\centering
  \includegraphics[width=.58\linewidth]{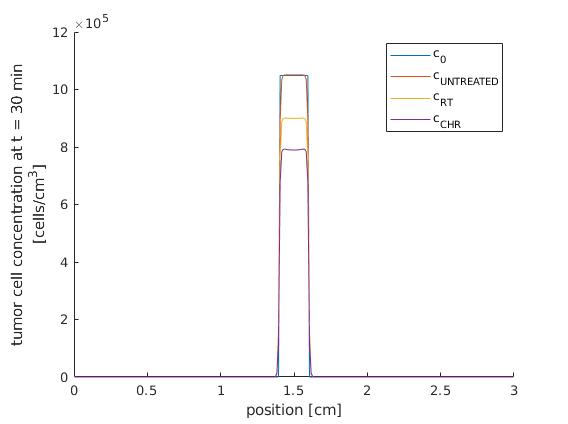}
   \caption{Tumor cell concentration comparison}
  \label{fig:fig1}
\end{figure}

\Cref{fig:fig1} shows that the introduction of radiation in the tumor begins to kill cancer cells and reduces the total tumor cell concentration in the slab after thirty minutes of treatment. 
Introduction of heat enhances the cell kill; this is evident from the further reduction in tumor cell concentration.\vspace{-5pt}

\begin{figure}[!h]
\centering
  \includegraphics[width=.58\linewidth]{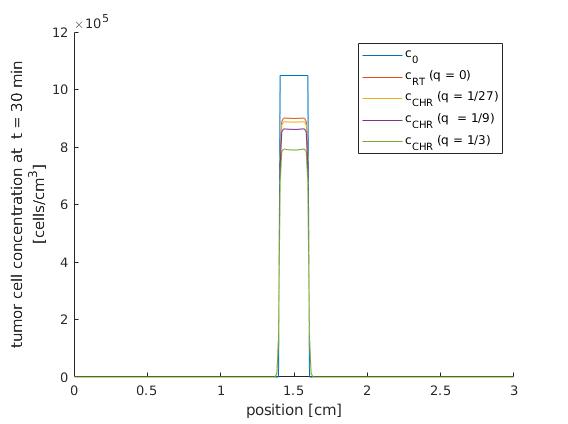}
   \caption{Tumor response to different heat sources}
  \label{fig:fig2}
\end{figure}

Finally, we compare tumor response to CHR treatment with different heat sources.
We vary the strength of the volumetric heat source in the tumor region of the model slab: $q = \frac{1}{27}$, $\frac{1}{9}$, and $\frac{1}{3}$ $\frac{W}{cm^3}$, and we fix $Q_0$ at $5.6\times10^{5}$ $\frac{photons}{cm^3 s}$. 
We find from \cref{fig:fig2} that increasing the strength of the heat source indeed results in a higher tumor cell kill. 
This is attributed to increased radiosensitivity in tumor tissue due to the introduction of hyperthermia.

\section{CONCLUSIONS}\label{sec4}

We presented a one-way coupled multiphysics model to determine tumor response to CHR treatment. 
We observed that the introduction of radiation results in tumor cell kill.
Moreover, introduction of heat enhances radiosensitivity of the tumor cells, which increases the cell kill. 
We made several simplifying assumptions and chose a simple geometry to demonstrate the concept behind this multiphysics model. 
We plan to eliminate these simplifying assumptions in subsequent papers such that real-world treatments can be modeled.
Specifically, we will introduce anisotropic scattering, higher dimensions, and energy dependence into our transport model. 
This will call for an efficient nonlinear HOLO scheme for forward-peaked transport problems; we will endeavour to develop such a method.
We also plan to extend this model to address random tumor media.   
Moreover, we understand the importance of developing a benchmark to verify future models and code implementations. 
This will constitute a major portion of our future work in CHR modeling.
\vspace{-5pt}

\section*{ACKNOWLEDGMENTS}

J.~K.~Patel and R.~Vasques acknowledge support under award number NRC-HQ-84-15-G-0024 from the Nuclear Regulatory Commission.
The statements, findings, conclusions, and recommendations are those of the authors and do not necessarily reflect the view of the U.S. Nuclear Regulatory Commission.
\vspace{-5pt}

 \bibliographystyle{mandc}
 \bibliography{mandc}

%

\end{document}